\documentclass[prl,reprint,showpacs,superscriptaddress,twocolumn,aps]{revtex4-1}

\usepackage{graphicx}
\usepackage{dcolumn}
\usepackage{bm}

\usepackage{mathtools}
\usepackage{color}
\usepackage{graphicx}
\usepackage{dcolumn}
\usepackage{amsbsy}
\usepackage{amsfonts}
\usepackage{color,soul}
\usepackage{amssymb}
\usepackage{enumitem}
\usepackage{bm}

\newcommand{\be}{\begin{equation}}
\newcommand{\ee}{\end{equation}}
\newcommand{\bea}{\begin{eqnarray}}
\newcommand{\eea}{\end{eqnarray}}

\newcommand{\Tr}{{\rm Tr\,}}

\newcommand{\<}{\langle}
\renewcommand{\>}{\rangle}

\renewcommand{\vec}[1]{{\bf #1}}
\renewcommand{\phi}{\varphi}
\renewcommand{\epsilon}{\varepsilon}

\newcommand{\ve}{\varepsilon}

\begin{document}

\title{Stability of Periodically Driven Topological Phases against Disorder}

\author{Oles Shtanko}
\affiliation{Physics Department, Massachusetts Institute of Technology, Cambridge, Massachusetts 02139, USA}
\author{Ramis Movassagh}
\affiliation{IBM Research, MIT-IBM A.I. Lab, 75 Binney Street, Cambridge, MA 02142}

\begin{abstract}
In recent experiments, time-dependent periodic fields are used to create exotic topological phases of matter with potential applications ranging from quantum transport to quantum computing. These nonequilibrium states, at high driving frequencies, exhibit the quintessential robustness against local disorder similar to equilibrium topological phases. However, proving the existence of such topological phases in a general setting is an open problem. We propose a universal effective theory that leverages on modern free probability theory and ideas in random matrices to analytically predict the existence of the topological phase for finite driving frequencies and across a range of disorder. We find that, depending on the strength of disorder, such systems may be topological or trivial and that there is a transition between the two. In particular, the theory predicts the critical point for the transition between the two phases and provides the critical exponents. We corroborate our results by comparing them to exact diagonalizations for driven-disordered 1D Kitaev chain and 2D Bernevig-Hughes-Zhang models and find excellent agreement. This Letter may guide the experimental efforts for exploring topological phases.
\end{abstract}

\maketitle 

The dynamics of nonequilibrium quantum systems has
been a subject of active and recent study with experiments involving several dozens of qubits \cite{bernien2017probing,zhang2017observation}. A promising
technique for creating nonconventional states of matter is by the application of a time-periodic field (e.g., to interacting cold atoms). These nonequilibrium states of matter are frequently referred to as $\,$\textit{Floquet} phases \cite{bukov2015universal,eckardt2017colloquium}.  The propositions and realizations include Floquet topological insulators 
 \cite{Kitagawa2012,hauke2012non,atala2013direct,wang2013observation,meier2016observation,cayssol2013floquet}, anomalous Floquet-Anderson insulators \cite{titum2016anomalous,nathan2017stability,maczewsky2017observation}, discrete time crystals \cite{else2016floquetTC,choi2017observation}
 etc. Remarkably, the
controlled periodic driving helps create Majorana modes
with non-Abelian braiding statistics potentially useful in
topological quantum computation \cite{jiang2011majorana,liu2013floquet,kundu_2013}. 

Local disorder is inevitable in realizing such nonequilibrium phases. Yet engineered systems can utilize artificial
disorder as a tool for control \cite{nathan2017stability,choi2017observation}. For example, disorder leads to many-body localization \cite{bordia2017periodically,abanin2018ergodicity}  preventing uncontrolled heating \cite{ponte2015periodically,else2016floquetTC} and stabilizing topological phases of matter \cite{vonKeyserlingk_2016,chen_2015,else_2016,bahri2015localization}. Disorder is also responsible for phase
transitions \cite{titum2015disorder,titum_2017,khemani_2016,gannot2015effects,pikulin2014disorder}. Even though topological phases in
equilibrium are universally robust against disorder, their
Floquet counterparts may not be. In low-dimensional
systems, the stability is typically granted by the
Anderson localization preserving the bulk mobility gap,
even if the disorder closes the bulk spectral gap
\cite{li_2009,groth_2009}. The same mechanism protects Floquet topological phases
at high frequencies \cite{Keyserlingk_2016}. However, if the driving frequency
is finite, Anderson localization may break down depending
on the driving amplitude and disorder strength \cite{Mott_1970,Klein2007,ducatez2017anderson, Agarwal_2017}. In this regime, nothing can preserve the topological phase if the bulk spectral gap is closed by disorder.

Despite the numerical frontiers  \cite{Keyserlingk_2016,titum2015disorder,titum_2017}, it is very difficult to quantify disordered Floquet systems in general. Even though in the limits of high driving frequency and weak disorder one can use techniques such as perturbation theory, many current realizations operate outside these limits \cite{jiang2011majorana,liu2013floquet,else2016floquetTC}. This raises the following questions:
Are Floquet topological phases preserved under finite
frequency and strong disorder? And if there is a disorder-induced transition into a trivial phase, can one quantify the critical point in the thermodynamic limit?
\begin{figure}[!t]
\begin{center}
 \includegraphics[width=\columnwidth]{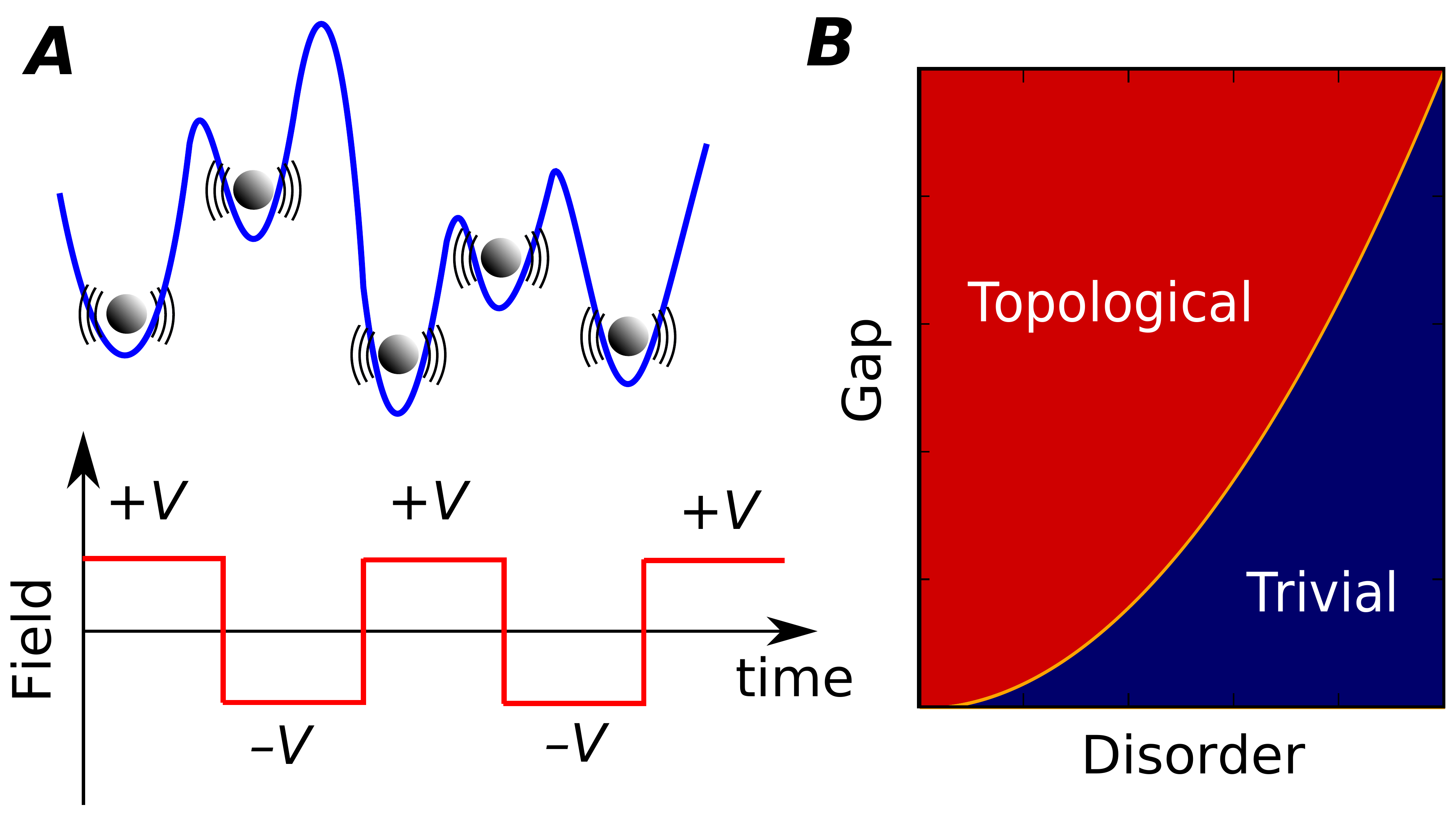}
\caption{\textbf{Schematics.}
\textbf{A} An isolated disordered quantum system represented by trapped cold atoms. The time periodic field $V(t)$ induces a transition from a trivial to a topological phase. \textbf{B} The phase diagram for the system in the presence of local disorder. An increase of the disorder strength $W$ induces a phase transition at $W_c\sim \Delta_0^{1/2}$, where $\Delta_0$ is the gap of the clean system.
}\label{fig:schematics} \end{center}\end{figure}

\begin{figure}[!t]
\begin{center}
 \includegraphics[width=\columnwidth]{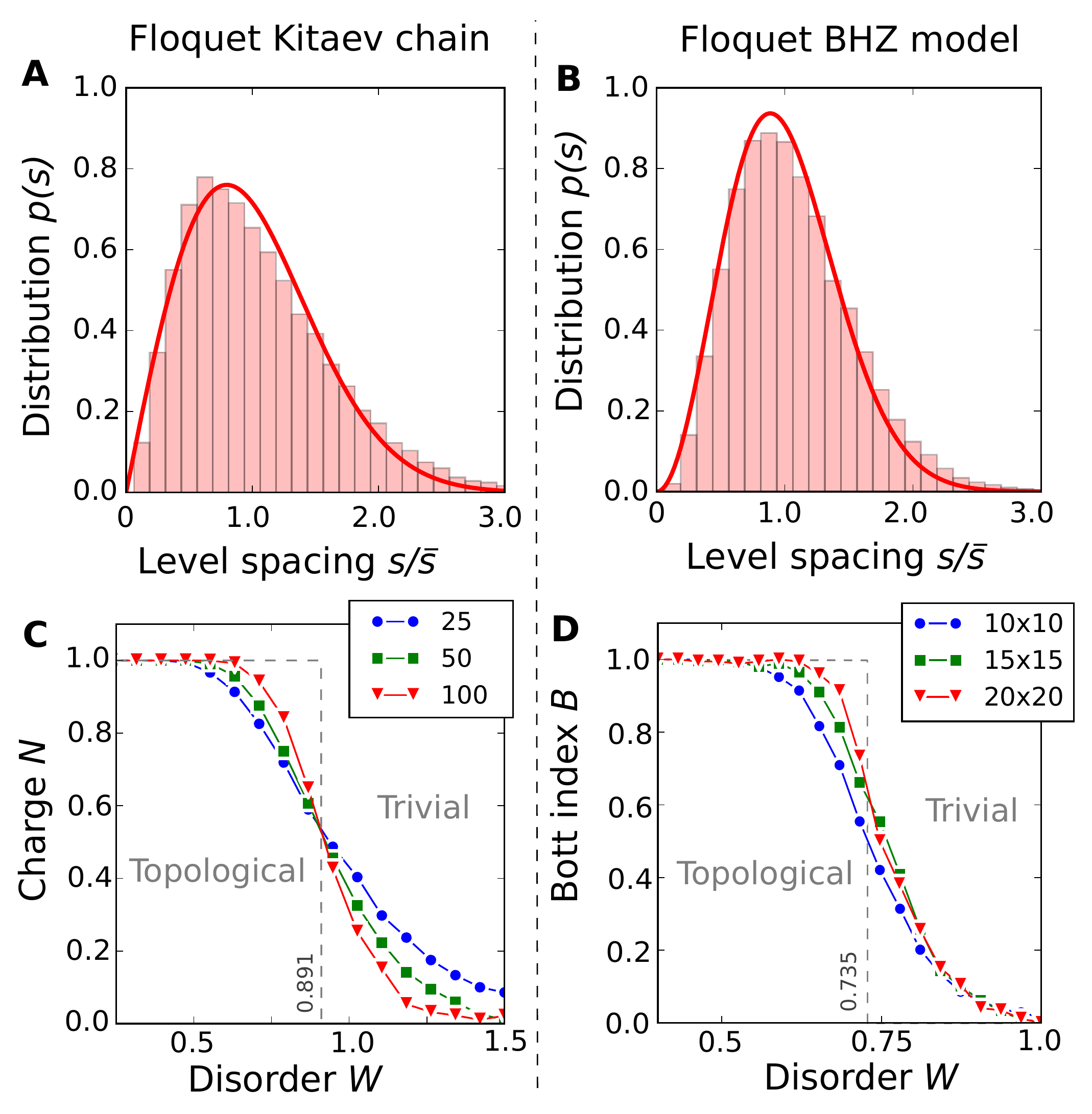}
\caption{\textbf{Effect of disorder}.
\textbf{A}, \textbf{B} Level spacing distribution for the middle of the spectrum of $\delta V_F$ for Floquet Kitaev chain and Floquet BHZ models, respectively, for $W=0.5$. Red curves are the level spacing distribution for GOE (\textit{A}) and GUE (\textit{B}), respectively.
\textbf{C}, \textbf{D} The topological charge and Bott index as a function of the disorder for the Kitaev chain and BHZ models, respectively. The dashed step function is the expected behavior in the thermodynamic limit. The parameters of the models are as in Fig.$\,$\ref{Fig:DOS_Gap}.}
\label{fig:EffectDisorder}
\end{center}
\end{figure}

\begin{figure}[t!] 
\includegraphics[width=1\columnwidth]{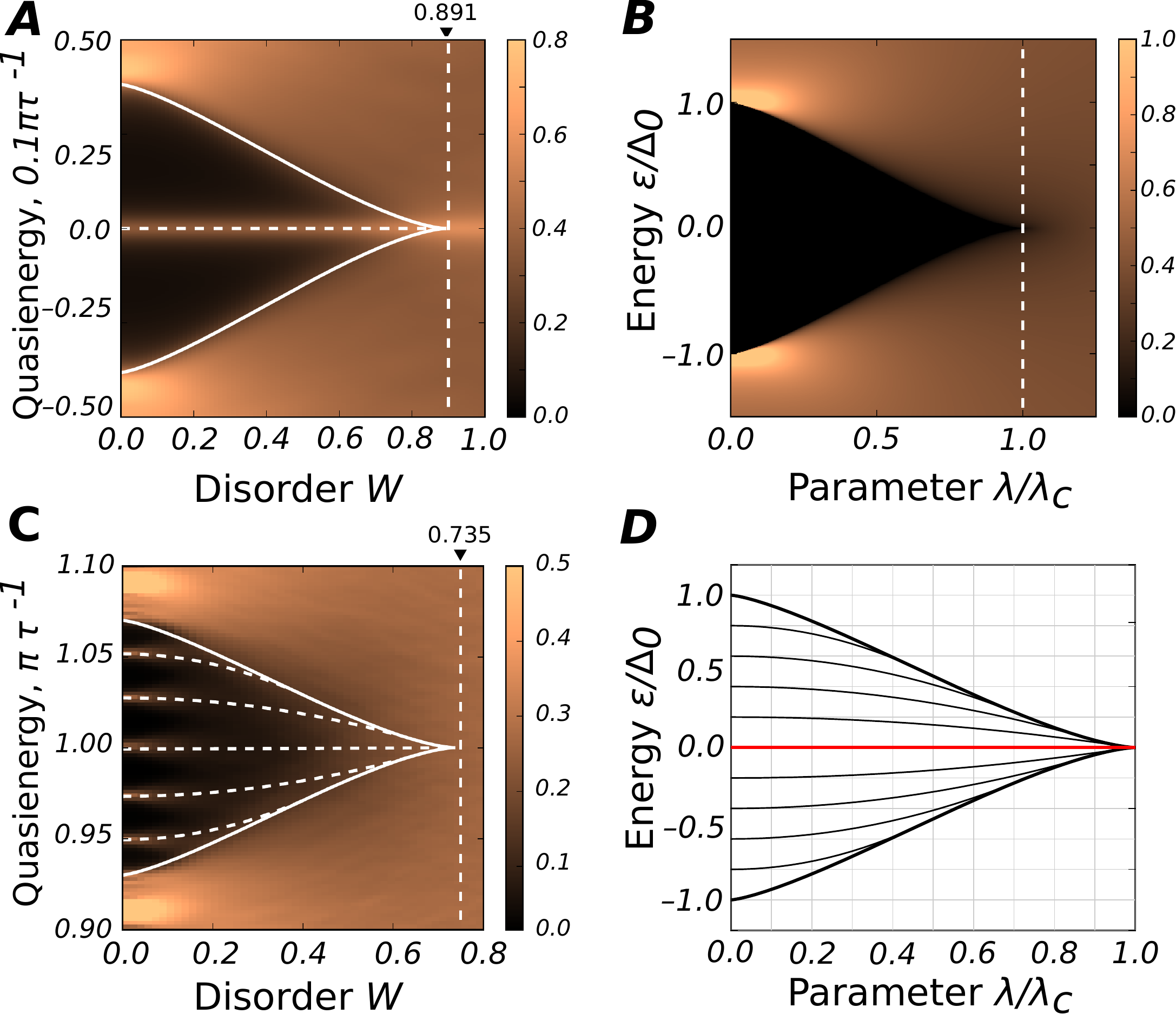}
\caption{\textbf{Density of states (DOS).} \textbf{A} DOS of Floquet 1D Kitaev chain of size $L=10^2$ for $\Delta=1$, $J=1$ $\mu=4.5$, $f=1.5$ and $\tau = 2\pi/\Omega= 1.1$ (level width applied $\gamma=10^{-2}\tau^{-1}$).  White solid curve and dashed white line are the analytical gap prediction (Eq.\eqref{eq:gap_behaviour}) with $\lambda = W$ and the Majorana state, respectively.
\textbf{B} DOS of Floquet BHZ model on a square lattice $20\times20$ and mixed periodic ($x$-direction) and open ($y$-direction) boundary conditions near quasienergy $\ve = \Omega/2$. (Level widening applied $\gamma = 0.2\, 10^{-2}\tau^{-1}$). The white curve is the analytical gap prediction given by Eq.\eqref{eq:gap_behaviour} with $\lambda = 0.9 W$, and the dashed curves are the analytical prediction for behavior of the mid-gap states given by Eq.\eqref{eq:midgap_states}. 
\textbf{C} Analytic calculations of bulk DOS for the model Eq.\eqref{eq:order_to_chaos_ham}, as described below Eq. \eqref{eq:inv_GF_total}. \textbf{D} Analytic result for the mid-gap states given by Eq.\eqref{eq:midgap_states}.}
\label{Fig:DOS_Gap}
\end{figure}

In this Letter, we leverage on modern free probability
theory and ideas inspired by random matrices to answer
these questions. The local disorder in the Hamiltonian
introduces a correction to the Floquet Hamiltonian
(Eq. \eqref{eq:delta_VF}). At finite driving frequencies, this correction is the sum of (potentially infinitely) many noncommuting terms in the Magnus expansion. Due to its nonlocality and randomness, we find that this correction has level statistics very similar to the Gaussian orthogonal (GOE) or unitary (GUE) ensembles depending on the problem (Figs.$\,$\ref{fig:EffectDisorder}$\,$A$\,$and$\,$B). We propose an effective model for the disordered
Floquet Hamiltonian, in which the correction is replaced by
a single generic random matrix proportional to the strength
of disorder (Eq.$\,$\eqref{eq:order_to_chaos_ham}). We use free probability theory
to analytically demonstrate that the effective Floquet
Hamiltonian does indeed exhibit a topological phase at a
finite strength of disorder and finite driving frequency. We
also find a critical strength of disorder beyond which the
spectral gap closes. Consequently, a transition is induced
from a topological into a trivial metallic phase. The
resulting phase diagram is shown in Fig.$\,$\ref{fig:schematics}B. We compare the universal analytical results against exact diagonalization for the disordered Kitaev chain and the 2D Bernevig-
Hughes-Zhang (BHZ) model (see Fig.$\,$\ref{Fig:DOS_Gap}).

Consider the problem of noninteracting particles on a
lattice. It is useful to divide the Hamiltonian into three
parts: the translationally invariant static part $H_0$, the static local disorder $\delta V$, and the applied external time-periodic driving field $V(t)$ (Fig.$\,$\ref{fig:schematics}A). Therefore,
\be\label{eq:Ht}
H(t) = H_0+\delta V + V(t), \qquad V(t)=V(t+\tau),
\ee
where $\tau$ is the driving period. Since topological phases in the integer quantum Hall universality class are often
understood in terms of free particles, we leave the effects
of many-body interactions for future work.

Let us first focus on the clean system. By the Floquet-Bloch theorem, the total time evolution at discrete times $t=n\tau$ is given by the unitary operator $U_n = (U_F)^n$, where  $U_F\equiv\exp (-i \tau H_F) =\mathcal T \exp\Bigl(-i \int_0^\tau dt' H(t')\Bigl)
$, $\mathcal T$ denotes chronological time ordering, and $H_F$ is the Floquet Hamiltonian. 
The Hamiltonian $H_F$ defines a new quantum (Floquet) phase \cite{vonKeyserlingk_2016}. Depending on the field $V(t)$, this phase can be equivalent to the initial phase of $H_0$, or be different.
Here we focus on the latter case where the field $V(t)$ is designed to convert a trivial into a topological phase \cite{bukov2015universal,kuwahara2016floquet}. 

Next we look at the role of disorder, $\delta V$, which may be represented by a diagonal random matrix. Periodic driving $V(t)$ and $\delta V$ dress the bare Floquet Hamiltonian into a disordered Floquet Hamiltonian $H'_F$ defined by
\be\label{eq:delta_VF}
H'_F = H_F + \delta V_F ,
\ee
where $\delta V_F\equiv \sum_{\ell \ge 1} \delta V_\ell \text{ } \tau^{\ell-1}$ with the coefficients
\be\label{eq:Magnus_expansion}
\delta V_1 = \delta V, \quad \delta V_\ell = \frac1{\tau^\ell}\Bigl[K_{\ell}\{H(t)+\delta V\}-K_{\ell}\{H(t)\}\Bigl],
\ee
denoting by $\{ .\}$ a functional. $K_\ell$ is the $\ell$th term in the Magnus expansion (Ref. \cite{bukov2015universal} and the Supplemental Material). In contrast to the random on-site potential $\delta V$, in
principle, each $\delta V_\ell$ contributes nonzero off-diagonal entries to the matrix $\delta V_F$, making the effective disorder nonlocal.

If the high driving frequency limit $\Omega =2\pi/\tau\rightarrow\infty$, the
higher-order corrections can be neglected. As a result,  $\delta V_F$ acts similarly to the local disorder $\delta V$, leading to the localization of eigenstates. In this situation, $H_F+\delta V_F$ always describes the topological phase as soon as a
mobility gap is present in the system.

On the other hand, if $\Omega$ is finite, the higher order terms in Eq.\eqref{eq:Magnus_expansion} cannot be ignored as $\tau$ may exceed the radius of convergence of Eq.\eqref{eq:Magnus_expansion}. Consequently, the off-diagonal entries in $\delta V_F$ are not negligible. Physically, this corresponds to emergence of driving-induced Landau-Zener transitions between localized states responsible for the breakdown of Anderson localization. In this regime, if the spectral gap closes, the Floquet topological phase is breaking with following disorder-induced transition to a
trivial phase.

In general, obtaining the exact spectral properties of
Eq.$\,$\eqref{eq:delta_VF} analytically is formidably difficult, mainly limited by the noncommutativity. Further numerical simulations are limited for large system sizes. However, there are many nondiagonal corrections appearing in Eq.$\,$\eqref{eq:Magnus_expansion}; the disorder
$\delta V$ added to $H(t)$ smears all over the effective Floquet
Hamiltonian (i.e., $\delta V_F$ in Eq.$\,$\eqref{eq:delta_VF}). It then seems plausible to
assume that the resulting $\delta V_F$ should mimic a generic
Hermitian random matrix. Indeed, in Figs.$\,$\ref{fig:EffectDisorder} A and B,
we show the accuracy by which the level statistics of $\delta V_F$
are represented by the standard Gaussian ensembles. We
will return to this below.

Therefore, the {\it effective Hamiltonian} we propose is:
\be\label{eq:order_to_chaos_ham}
H^{\rm eff}_F = H_F + \lambda\, M,
\ee
where the matrix $M$ is chosen from the Gaussian ensemble
with eigenvalues in $[-2,2]$, which in the limit of infinite
size would follow the semicircle law \cite{mehta2004random}, and $\lambda = \sqrt{\varphi(\delta V_F^2)}$
with $\varphi(A) = \mathbb{E}\Tr (A)/{\rm dim}(A)$ denoting the empirical mean of the matrix $A$. Physically, $H^{\rm eff}_F$ describes
a competition between the topological phase ($\lambda<\lambda_c$) and featureless chaotic phase ($\lambda>\lambda_c$), where $\lambda_c$ is a critical
point. This model describes the transitions in a finite range of disorder strength and may not retain the precision in the limit of high disorder $W\gg\Omega$ in which the target Floquet Hamiltonian is expected to exhibit Poissonian quasienergy level statistics.

The value of the parameter $\lambda$ depends on both disorder strength $W$ and driving period $\tau$  (see Supplemental Material for details). In the weak disorder limit $\lambda \approx \alpha(\tau) W/\sqrt{3}$, where $\alpha(\tau)$ is a dimensionless parameter. At high driving frequencies $\alpha(\tau) = 1 + O(\tau^2)$. This approximation is valid if the period of driving is less than the radius of convergence of Eq.$\,$\eqref{eq:Magnus_expansion} \footnote{Radius of convergence is given by $\int_0^\tau ||H(t)||dt\leq \pi$, see Supplementary Material and \cite{blanes2009magnus}}. At low frequencies, $\lim_{\tau\to \infty}\alpha(\tau) = \alpha_0$, where $\alpha_0$ is a constant that depends on the model. In the strong disorder limit, assuming that the eigenvalues of $\delta V_F\tau$ are evenly distributed in the interval $[-\pi,\pi]$, we get $\lambda\tau \approx \pi/\sqrt{3}$. The value of $\lambda$ plays the role of a phenomenological parameter in the model.

To this end, and before presenting the analytical machinery, we demonstrate our results in the context of two widely studied models, the Kitaev chain and the 2D Bernevig-Hughes-Zhang (BHZ) model. For numerical simulations, both models can be represented as particular cases of a fermion hopping on a lattice:
\be
\begin{split}
\label{eq:general_ham}
H_0 = \sum_{\vec r,\vec a\in A}\Bigl(\Gamma_{\vec a} |\vec r\>\<\vec r+\vec a|+{\rm h.c.}\Bigl)+\mu \Gamma_0,
\end{split}
\ee
where $A = \{\vec a\}$ is the set of primitive vectors on the lattice,
$\Gamma_{\bf a}$ and $\Gamma_0$ are Hermitian matrices, and $\mu$ is the chemical
potential. We choose the driving field and disorder to be
\be
V(t) = F\theta(t),\qquad \delta V = \Gamma_0\sum_{\vec r} h_{\vec r}|\vec r\>\<\vec r|,
\ee
where $\theta(t) = {\rm sgn}(\sin \Omega t)$, and $h_{\bf{r}}$ is uniformly random in $[-W,W]$. 

As the first example, we take the Bogoliubov–de Gennes
Hamiltonian for the Kitaev chain \cite{kitaev2001unpaired}, which has the form
of Eq.$\,$\eqref{eq:general_ham} with  $\Gamma_x = i\Delta \sigma_y + J\sigma_z$, $\Gamma_0 = \sigma_z$, and $F = f\sigma_z$.
The $\sigma_i$’s are the Pauli matrices, $\Delta$  is the superconducting
pairing, $J>0$ is the hopping constant, and $f$ is the
amplitude of the external driving. In the absence of driving,
the clean system is an archetypal example of a 1D
topological superconductor, exhibiting a topological phase
transition at $|\mu|=2J$. When $|\mu|<2J$, the system is in the
topological phase hosting two Majorana zero-energy
modes at each end of the chain, and is in the trivial phase
otherwise. However, recent proposals \cite{jiang2011majorana,liu2013floquet} suggest that
when $|\mu|>2J$, the Kitaev chain may also exhibit topological
states if a local periodic field is applied  ($f\neq0$). In
this case, Majorana modes can exist for quasienergies $\ve =0$ and $\ve=\Omega/2$. We focus on the stability of the $\ve =0$
Majorana mode against disorder present in the system,
and we assume similar behavior for the $\ve=\Omega/2$ mode.

Numerically, we find that strong disorder destroys the induced topological Floquet phase by closing the spectral gap. Fig.$\,$\ref{fig:EffectDisorder}C demonstrates this transition for the average topological charge $N=-\<{\rm sign}({\rm Pf}(i\tilde{H}_F))\>_{\rm dis}$, where $\tilde{H}_F$ is the Floquet Hamiltonian in the Majorana representation (it is Hermitian and purely imaginary), and ${\rm Pf}$ is Pfaffian. If $N=1$, the system is in a topological phase, while in the disordered trivial phase $N=0$. Vanishing of the gap corresponds to the transition from $N=1$ to $N=0$. Fig.$\,$\ref{Fig:DOS_Gap}A shows the closing of the gap at the critical disorder strength. The analytical predictions of the effective theory Eq.\eqref{eq:order_to_chaos_ham} with $\lambda=W$ are also presented in Fig.$\,$\ref{Fig:DOS_Gap}A. The analytical calculation of the gap (white solid curve) and zero-energy mode (white dashed line) show good agreements with exact diagonalization.

Similar results can be obtained in 2D by choosing a
square lattice with $\Gamma_x = -i\frac{A}2 \sigma_x + B\sigma_z$, $\Gamma_y = -i\frac{A}2 \sigma_y + B\sigma_z$, and 
$M_{\vec r} = h_{\bf r}+(\Delta-4B+f{\rm sgn}(\sin \Omega t))\sigma_z$. Here $A$ is
a velocity parameter, $B$ defines the inverse kinetic mass,
$\Delta$ is the bulk band gap, and $h_{\bf r}$ is the static disorder. The
long-wavelength limit of this model coincides with the
seminal BHZ theory \cite{bernevig2006quantum} $H_0 = \sum_{i=1}^3 d_i(\vec k)\sigma_i$, where $d(\vec k) = (Ak_x,Ak_y,\Delta-B(k_x^2+k_y^2))$. For $\Delta/B>0$, the disordered
system is characterized by the Bott index \cite{loring2011disordered,toniolo2017equivalence}
$C=1$ and hosts topologically protected states at the edge.
Similar to the Kitaev chain, the trivial phase $\Delta/B<0$ can
be converted into a topological phase by applying a
periodic driving field $f\neq0$. The effect of disorder is
shown in Fig.$\,$\ref{fig:EffectDisorder}D and Fig.$\,$\ref{Fig:DOS_Gap}C. In Fig. \ref{Fig:DOS_Gap}C, the white solid
curve and dashed white curves are the gap and edge states,
respectively; both are analytically computed. The discreteness
of edge states is due to the finite size.

The efficacy of $H^{\rm eff}_F$  (Eq.$\,$\eqref{eq:order_to_chaos_ham}) in capturing the exact $H'_F$ (Eq. \eqref{eq:delta_VF}) is easily demonstrated in the models we studied by examining the matrix $\delta V_F = H_F-H_F\bigl|_{W=0}$. Remarkably,
$\delta V_F$ turns out to be a nonlocal matrix with level statistics
close to the Wigner-Dyson law (Fig.$\,$\ref{fig:EffectDisorder}A-B).
Interestingly, the renormalized disorder $\delta V_F$  approximately
follows the GOE and GUE statistics for the 1D Kitaev
chain and 2D BHZ Hamiltonian, respectively. Whether
GOE or GUE level statistics is dictated by the dimension of
the lattice is a question we leave for future work.

We proceed with our main goal of analytically solving
spectral properties of $H^{\rm eff}_F$ (Eq.$\,$\eqref{eq:order_to_chaos_ham}). The main tool enabling
this is free probability theory \cite{mingo2010free,movassagh2011density,chen2012error,movassagh2017eigenvalue}, which we now
introduce (see the Supplemental Material and Ref. \cite{movassagh2017eigenvalue}
for an applied overview). Free probability theory (FPT)
extends the conventional probability theory to the noncommuting
random variables setting. Recall the $\varphi$ notation $\varphi(A) = (\mathbb{E}\Tr A)/{\rm dim}(A)$, and $\overline {A^k} = A^k-\varphi(A^k)$. Two random
matrices $A$ and $B$ are freely independent (or free) if all
expectation values of cross-term correlators vanish in the
infinite size limit. That is,That is $\varphi(\overline {A^{k_1}}\,\overline {B^{l_1}}\,\dots \overline {A^{k_n}}\,\overline {B^{l_n}})=0$ (see Refs. \cite{nica2006lectures,movassagh2017eigenvalue} for a comprehensive
definition). The free independence is immediate if either A or B is chosen independently from the Gaussian ensemble.
Therefore, in Eq.$\,$\eqref{eq:order_to_chaos_ham}, $H_F$ and $\lambda M$ are free.

The input to the theory is the Cauchy transform of the
DOS of the summands $G_A(z)=\varphi( (z-A)^{-1})$ and $G_B(z)=\varphi( (z-B)^{-1})$. The integral representation of $\rho_A(\ve)$, of matrix $A$  is (similarly for $B$) is
\be\label{eq:cauchy_transform}
G_A(z) = \frac1{2\pi i}\int_{\mathbb{R}} d\ve\text{ } \frac{\rho_A(\ve)}{z-\ve}.
\ee
The R transform is defined by $R_A(w) = G_A^{-1}(w)-w^{-1}$,
where $G_A^{-1}$ is the functional inverse (similarly for $B$). Recall
that in standard probability theory, the additive quantity for
sums of scalar random variables is the log characteristics. In
FPT, the analogous additive quantity is the R transform,
which in turn defines the Cauchy transform of the sum
$G_{A+B}(\ve)$. One then obtains the DOS from $G_{A+B}(\ve)$, with
the caveat that the technical challenge often is the inversion
of $G_{A+B}(\ve)$ to obtain the density. Below, $H_F$ and $\lambda M$ replace $A$ and $B$, respectively (see the Supplemental
Material for technical details of what follows).

The R transform of $H^{\rm eff}_F$ in Eq.\eqref{eq:order_to_chaos_ham} is $R_{H^{\rm eff}_F}(w)\equiv R_{H_F}(w) + R_{\lambda M}(w)$. This is equivalent to (see the
Supplemental Material)
\be\label{eq:additivity_G}
G^{-1}_{H_{\rm eff}}(w) = G^{-1}_{H_F}(w)+G^{-1}_{\lambda M}(w)-\frac{1}{w}.
\ee
At energies not much larger than the Floquet band
gap $\Delta_0$, the bulk DOS of the topological Hamiltonian
$H_F$ is approximated by $\rho_{H_F}(\ve) \approx \rho_0\ve/\sqrt{\ve^2-\Delta_0^2}$, where
$\rho_0$ is the DOS in the vicinity of the gap. The DOS
of $\lambda M$ is the well-known semicircle law $\rho_{\lambda M}(\ve) = (2\pi\lambda^2)^{-1}\sqrt{4\lambda^2-\ve^2}$. The Cauchy transform $w\equiv G_{H_{\rm eff}}(\ve)$ can be derived from the condition Eq.$\,$\eqref{eq:additivity_G}, which is
equivalent to
\be\label{eq:inv_GF_total}
\ve = \lambda^2 w+\frac{w^2\Delta_0^2}{\sqrt{\pi^2\rho_0^2-w^2}}.
\ee
The DOS is then obtained from the imaginary part of the
Cauchy transform, $\rho(\ve) = \pi^{-1}{\rm Im\,}w$. Energies $\ve$ for which $w$
is real in Eq.$\,$\eqref{eq:inv_GF_total} correspond to zero density of states -- i.e.,
the band gap. The real solutions of $w$ occur for $\lambda<\lambda_c$, with
\be\label{eq:critical_lambda}
\lambda_c = \sqrt{\Delta_0/\pi\rho_0}.
\ee
$\lambda_c$  defines the critical point for the phase transition, where
two bands merge and the gap vanishes (Fig. \ref{Fig:DOS_Gap}B). Let $\Delta(\lambda)$
be the band gap as a function of the effective disorder
strength $\lambda$. For $|\ve|<\Delta$, one has
\be\label{eq:gap_behaviour}
\Delta(\lambda) = \Delta_0 \left[1-(\pi\rho_0\lambda^2/\Delta_0)^{2/3}\right]^{3/2},\quad \lambda<\lambda_{\rm c},
\ee
and $\Delta(\lambda)=0$ for $\lambda\ge\lambda_c$.

We turn our attention to the behavior of the surface states
energies $\ve_\mu$ situated in the bulk band gap, where $\mu$ can be
either a discrete or a continuous quantum number. To
evaluate $\ve_\mu(\lambda)$, one can use the fact that the number of
surface states is small compared to bulk ones. This allows
us to derive the spectrum, considering them as small
corrections to the Cauchy transform (Eq.\eqref{eq:cauchy_transform}).

In the Supplemental Material, we show that the
resulting spectrum of midgap states satisfies $G_{H_F}(\ve_\mu) = G_{H_{\rm eff}}(\ve_\mu(\lambda))$, which leads to
\be\label{eq:midgap_states}
\ve_\mu(\lambda) = \ve_\mu\Biggl(1-\frac{\pi\rho_0\lambda^2}{\sqrt{\Delta_0^2-\ve_\mu^2}}\Biggl), \qquad \lambda<\lambda^\mu_{\rm c},
\ee
where $\lambda_c^\mu$ denotes the solution of $\ve_\mu(\lambda_c^\mu) = \Delta(\lambda_c^\mu)$. The plots for $\ve_\mu(\lambda)$ for different initial values $\ve_\mu$ are shown in Fig.$\,$ \ref{Fig:DOS_Gap}D. As seen there, the continuous spectrum of surface states never opens up a spectral gap.

{\it Remark}. -- The theory is universal, in that the details of
the underlying model, such as the dimension of the lattice,
the period $\tau$, or the DOS of the clean system, only enter
through $\rho_0$ and $\Delta_0$.

To summarize, we demonstrated that the disorder effects
on finite-frequency Floquet phases can be well approximated
by generic random matrices (Eq. \eqref{eq:order_to_chaos_ham}). Using this and
free probability theory, we analytically show that the
topological phases in this regime are generically stable
against disorder for a range of strength. The breakdown into
the trivial phase typically happens at a critical disorder
strength that is potentially many times larger than the
spectral gap. The proposed theory allows us to compute the
critical gap behavior and the corresponding critical exponents.
The analytical prediction of the critical point can
serve as a guide in experiments to search for topological
phases in the presence of disorder more systematically and
irrespective of the underlying model.

The utility of free probability theory for approximating
spectral properties of physical systems extends beyond this
Letter. On the one hand, it works in the more general
settings in which perturbative analysis fails (e.g., in the
current study, the regime of strong disorder and/or moderate
frequency of driving). On the other hand, free
convolution is an entirely new technique that can be added
to the arsenal of the existing tools. We emphasize that the
success of free probability theory does not rely on the
disorder being generic (cf. Refs. \cite{chen2012error, movassagh2011density}).

Future research may include applying our techniques to
time crystals \cite{berdanier2018floquet} and other disordered systems, especially
with many-body interactions—for example, the treatment
of the self-energy in self-consistent Born approximations
\cite{abrikosov1960contribution}. We anticipate these methods to provide a new angle of attack on problems of disordered superconductivity and
many-body localization.

\vspace{1em}

 We thank Iman Marvian. O. S. was supported by ExxonMobil-MIT Energy Initiative Fellowship. O. S. acknowledges MIT Externship program and partial support by IBM
Research.

\bibliography{bibliography}

\clearpage
\onecolumngrid

\begin{center}
\Large{\textbf{Supplemental Material}}
\end{center}
\vspace{0.5em}
\setcounter{page}{1}
\setcounter{equation}{0}
\setcounter{figure}{0}
\renewcommand{\theequation}{S.\arabic{equation}}
\renewcommand{\thefigure}{S\arabic{figure}}
\renewcommand*{\thepage}{S\arabic{page}}


\section{Calculation of spectrum using free probability theory}
Free probability theory (FPT) extends the conventional probability theory to the setting in which the random variables do not commute [1, 2]. 
The canonical examples of such random variables are random matrices. Since its discovery in 1980's, FPT has been mainly a sub-field of pure mathematics. However, in recent times, it has been distilled for applications and shown to have potentials for a wide set of problems of applied interest (see [3] for details and an overview of {\it applied} FPT). 

Suppose we are interested in the density of states (eigenvalue distribution) of the sum
\be
A = A_1 + A_2,
\ee
where the densities of $A_1$ and $A_2$ are known. If matrices are freely independent (see [1] for exact definition), FPT provides the distribution $\rho(\ve)$ of matrix $A$ from the densities $\rho_1(\ve)$ and $\rho_2(\ve)$ of matrices $A_1$ and $A_2$. The input to the theory is the {\it Cauchy transofrm} of the densities of the summands. The Cauchy transform of $\rho_\alpha(\epsilon)$ is defined by
\be\label{eq:Cauchy_transform}
G_\alpha(z) = \frac{1}{2\pi i}\int_{-\infty}^{\infty} d\ve\text{ } \frac{\rho_\alpha(\ve)}{z-\ve}.
\ee 
It is good practice to introduce a new variable, $w$, to denote the Cauchy transform $w\equiv G(z)$. 

Analogous to log-characteristics in conventional probability theory, the key additive quantify in FPT is the R-transforms:
\be\label{eq:R_additivity}
R = R_1+R_2, \qquad R_\alpha(w) \equiv G_\alpha^{-1}(w)-\frac 1w, \qquad \alpha = 1,2
\ee
where $z=G^{-1}_\alpha(w)$ is the functional inverse.

Technically, computation of the density of states of matrix $A$ can be performed in four steps:
\begin{enumerate}
\item Input to the theory are the Cauchy transforms of the summands denoted by  $G_1(z)$ and $G_2(z)$, which one obtains using Eq.\eqref{eq:Cauchy_transform},
\item Computation of the functional inverse for Greens functions $G^{-1}_1(w)$ and $G^{-1}_2(w)$;
\item One then finds the inverse Cauchy transform for sum of matrices $w=G^{-1}(w)$ using formula Eq.\eqref{eq:R_additivity} ;
\item Then one obtains the Cauchy transform of the sum, $w=G(z)$, by inversion. Lastly the density is computed by
\be
\rho(\ve) = \frac{1}{\pi}\lim_{\eta^+}\{{\rm Im}\, (G(z))\},
\ee
where $\frac{1}{\pi}\lim_{\eta^+}\left[{\rm Im}\, (G(z))\right]$ means taking a limit from above to the branch cut of $G(z)$. 
\end{enumerate}

Steps 2 and 4 require computing the functional inverse of corresponding Greens functions. In the generic case, it would require a numerical computation. However, in some physically relevant cases, the inversion can be performed analytically, as it is shown below.
\subsection{Problem of Random Disorder Correction}
Here we apply this method to the sum of matrices Eq.(3) in the main text describing the effect of disorder on the clean system Hamiltonian $H_F$:
\be
H_F^{\rm eff} = H_F + \lambda \text{ } M
\ee
where $\lambda$ is the real parameter that quantifies the strength of disorder.

Let us consider a Floquet Hamiltonian $H_F$ describing a $d$-dimensional non-interacting topological system of linear size $L$ with $N = N_B+N_S$ system eigenstates, where $N_B\sim L^d$ is number of bulk states and  $N_S\sim L^{d-1}$ is number of surface states. 
The density of states of the clean system is simply the sum of the bulk and surface states densities
\be
\rho(\ve) = \rho_{B}(\ve)+\rho_{S}(\ve)
\ee
We assume that the band gap $\Delta_0$ in the system is negligible compared to the bandwidth $\Gamma$, $\Delta_0\ll \Gamma$. This allows one to neglect the energy dependence of the DOS outside the gap in systems with quadratic spectrum, including superconductors. In this case, the DOS near the gap has the universal and dimension-independent form:
\be\label{eq:DOS_bulk}
\rho_{B}(\ve) \approx \frac{\rho_0|\ve|}{\sqrt{\ve^2-\Delta_0^2}},\quad\Delta_0\leq|\ve|\ll \Gamma
\ee
where $\rho_0$  and $\Delta_0$ depend on the details of the model. 

We only consider the surface states which are inside the gap. Let $\ve_\mu$ represent the spectrum of surface states (discrete for $d=1$ or continuous for $d\geq2$), then the corresponding surface states contribution to DOS is
\be \label{eq:DOS_surf}
\rho_S(\ve) = \frac{\alpha}{N_S}\sum_\mu \delta(\ve-\ve_\mu),\quad|\ve|<\Delta_0
\ee
where $\alpha = N_S/N\sim 1/L$ is a small parameter.
Both approximated bulk density expression Eq.\eqref{eq:DOS_bulk} and surface density expression Eq.\eqref{eq:DOS_surf}  are dimension-independent and universal across many models.

The disorder contribution is modeled by a Hermitian generic random matrix whose distribution is the well-known semicircle law
\be
\rho_{\lambda M}(\ve) = \frac1{2\pi\lambda^2}\sqrt{4\lambda^2-\ve^2}.
\ee
Comment: The underlying gaussian ensemble may be GOE or GUE.

\begin{figure}[!t]
\begin{center}
\includegraphics[width=\columnwidth]{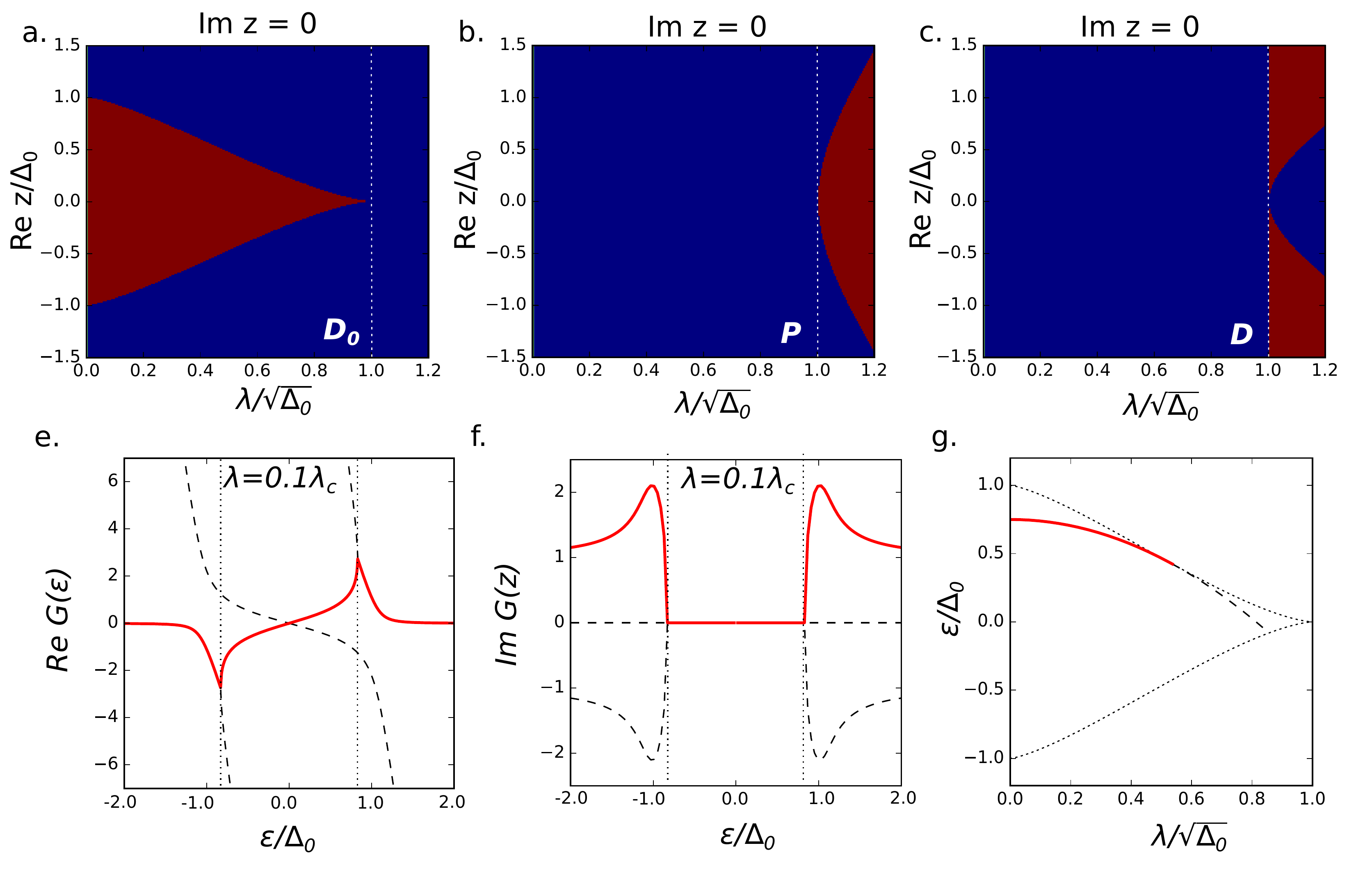}
\caption{\textbf{Details of analytical solution.} \textbf{a.-c.} Sign of discriminant $D_0(z,\Delta_0,\lambda)$ Eq.\eqref{eq:discriminant} and polynomials $P(z,\Delta_0,\lambda)$ and $D(z,\Delta_0,\lambda)$ Eq.\eqref{eq:PD_poly} as function of $z/\Delta_0$ taken at real $z$, and parameter $\lambda/\sqrt{\Delta_0}$. Region of positive parameter are red, regions with negative parameters are blue. The region where roots are purely real (region of spectral gap) coincides with the region of positive discriminant $D_0>0$. \textbf{e.-f.} Real and imaginary part of the solution for real $z$. Red lines represent the physically relevant solution of Eq.\eqref{eq:quartic_eq}. Black dashed lines represent other possible solutions. Black dotted lines show edges of the gap \textbf{g.} Midgap state solution. Red line denotes the physical solution for midgap state. Black dashed line represent the midgap solution due to non-physical solution for $G_H$.}
\label{fig:tf_plot}
\end{center}
\end{figure}

\subsection{Density of Bulk States}

Since $\alpha$ in Eq.\eqref{eq:DOS_surf} vanishes in the thermodynamic limit, one may ignore the influence of surface states on the bulk spectrum to estimate the behavior of the bulk states. The Cauchy transform 
is
\be\label{eq:Greens_HF}
G_{H_F}(z) = \frac1{N}\lim_{\eta\rightarrow 0}\Tr\frac1{z-H_F-i\eta} = \lim_{\eta\rightarrow 0}\int_{-\infty}^{\infty}d\ve\frac{\rho_B(\ve)}{z-(\ve+i\eta)} + O(\alpha) = \frac{z}{\sqrt{\Delta_0^2-z^2}}+O(\alpha)
\ee
where, to simplify the expressions, we drop $\pi\rho_0$ by rescaling 
$\lambda \rightarrow \pi\rho_0\lambda$, $z \rightarrow \pi\rho_0z$, and $\Delta_0 \rightarrow \pi\rho_0\Delta_0$.

The Cauchy transform of random matrix spectral density is
\be
G_{\lambda M}(z) = \lim_{\eta\rightarrow 0}\int_{-\infty}^{\infty}d\ve\frac{\rho_{\lambda M}(\ve)}{z-(\ve+i\eta)}  = \frac1{2\lambda^2}(z-\sqrt{z^2-4\lambda^2})
\ee
The R-Transform of the two distributions $\rho_B$ and $\rho_{\lambda M}$ are,  respectively, given by
\begin{align}
&R_{H_F}(w) = G^{-1}_{H_F}(w)-\frac1w = \frac{\Delta_0 w}{\sqrt{1+w^2}}-\frac1w\\
&R_{\lambda M}(w) = G^{-1}_{\lambda M}(w)-\frac1w = \lambda^2 w
\end{align}

 We now use the key additivity property of the R-transform, to obtain 
\be
R_{H_F^{\rm eff}}(w) = R_{H_F}(w)+R_{\lambda M}(w) = \lambda^2 w+\frac{\Delta_0 w}{\sqrt{1+w^2}}-\frac1{w}
\ee
With the R-transform of the sum in hand, we need to reverse obtain the actual density of the sum under the freeness assumption. The inverse Cauchy transform of the sum is $z\equiv G^{-1}_{H_F^{\rm eff}}(w)$
\be \label{eq:GF_inverse_full}
G^{-1}_{H_F^{\rm eff}}(w) = R_{H_F^{\rm eff}}(w)+\frac1{w} = w\Bigl(\lambda^2 +\frac{\Delta_0}{\sqrt{1+w^2}}\Bigl)
\ee
This equation needs to be inverted, and the inversion leads to solving the following polynomial equation
\be\label{eq:quartic_eq}
a\,w^4+b\,w^3+c\,w^2+d\,w+e=0,
\ee
where the coefficients are defined as
\be
a = \lambda^4,\quad b = -2z\lambda^2,\quad c = \lambda^4+z^2-\Delta_0^2,\quad d = -2z\lambda^2,\quad e = z^2.
\ee

Let us consider only real values of $z$. The discriminant of the polynomial is of the form
\begin{align}\label{eq:discriminant}
D_0(\ve,\Delta_0,\lambda) = 256 a^3 e^3-192 a^2 b d e^2-128 a^2 c^2 e^2+144 a^2 c d^2 e-27 a^2 d^4+144 a b^2 c e^2-6 a b^2 d^2 e \nonumber\\
-80 a b c^2 d e+18 a b c d^3+16 a c^4 e-4 a c^3 d^2-27 b^4 e^2+18 b^3 c d e-4 b^3 d^3-4 b^2 c^3 e+b^2 c^2 d^2\nonumber \\
=-16 \Delta_0^2 \lambda ^4 z^2 \left(z^{6}+3\left(\lambda^4-\Delta_0 ^2\right)z^4+3 \left(\Delta ^4 +7 \Delta_0 ^2 \lambda^4+\lambda^8\right)z^2+ \left(\lambda^4-\Delta_0 ^2\right)^3\right).
\end{align}
Two other quantities characterizing the quartic equation are
\begin{align}\label{eq:PD_poly}
&P(\ve,\Delta_0,\lambda)=8 a c-3 b^2 = -4 \lambda ^4 \left(2 \Delta_0^2+z^2-2 \lambda ^4\right)\\
&D(\ve,\Delta_0,\lambda)=64 a^3 e-16 a^2 b d-16 a^2 c^2+16 a b^2 c-3 b^4 = -16 \lambda ^8 \left(\Delta_0^2-\lambda ^4\right) \left(\Delta_0^2-\lambda ^4+2 z^2\right)
\end{align}
$\bf{Remark:}$ To have a gap, ones seeks the of parameters for which there is {\it no support} for DOS $\rho(\ve)$. Recall that
\be
\rho(\ve) = \pi^{-1}\lim_{\eta^+}\{{\rm Im}(\, G_{H_F^{\rm eff}}(z))\};
\ee
therefore we seek four real solutions to the quartic equation. This corresponds to having 
$P<0$, and $D<0$.

As can be seen from the analysis of the signs of $D_0$, $P$, and $D$, the solution of Eq.(\eqref{eq:quartic_eq}) has zero imaginary part only in the region where $D_0>0$ (see Fig. $\,$\ref{fig:tf_plot}). Therefore, the gap will be defined by the soultion of $D_0(\ve,\Delta_0,\lambda)=0$ (excluding solution $z=0$). In the explicit form it is equal to:

\be
z^{6}+3\left(\lambda^4-\Delta_0 ^2\right)z^4+3 \left(\Delta_0 ^4 +7 \Delta_0 ^2 \lambda^4+\lambda^8\right)z^2+ \left(\lambda^4-\Delta_0 ^2\right)^3=0.
\ee
Two real solutions of this equation $z = \pm z_0$ define the edges of spectral gap $\Delta(\lambda) = z_0$. It can be written in a compact form (here we restore the $\pi\rho_0$ factor we dropped starting at Eq.\eqref{eq:Greens_HF})
\be\label{eq:gap_behav}
\Delta(\lambda) = {\left(\Delta_0^{2/3} -(\pi\rho_0\lambda^2)^{2/3}\right)^{3/2}}.
\ee
As expected, at zero disorder $\Delta(0) = \Delta_0$ and decreases with disorder strength. At the critical strength $\lambda_c = \sqrt{\Delta_0/\pi\rho_0}$, the gap closes and the system transitions into the metallic phase. Using Taylor expansion, we obtain the behavior of the gap in the vicinity of the critical point:
\be
\Delta(\lambda) = \Delta_0\frac{8 }{3 \sqrt{3}}\frac{|\lambda-\lambda_c|^{3/2}}{\lambda_c^{3/2}}+O(|\lambda-\lambda_c|^{5/2}),\qquad \lambda<\lambda_c
\ee
From this one immediately reads off the critical exponents for such type of transition
\be\label{eq:critical_cond}
\nu z = 3/2
\ee
Since we neither specify exactly the Hamiltonian of the system nor its dimensionality, the condition Eq. \eqref{eq:critical_cond} is rather widely applicable to a variety of systems. \\

$\bf{Remark:}$ Eq. \eqref{eq:quartic_eq} is quartic and can be analytically solved. Nevertheless, obtaining the imaginary part of the solution, and consequently, DOS yields unwieldy expressions. Therefore, we obtain the solutions of Eq.\eqref{eq:quartic_eq} numerically  using \texttt{NumPy} Python package. The results are presented in Fig.$\,$3C of the main text.
\subsection{Density of Surface States}
The surface states give a contribution to the total DOS suppressed by a factor of $L^{-1}$. This enables one to calculate the corresponding DOS using perturbation theory. The Cauchy transform for the Floquet Hamiltonian $H_F$, including the surface states, reads as
\be\label{eq:full_Greensf}
\tilde G_{H_F}(z) = G_{H_F}(z) + \alpha \frac {2\pi}{N_S}\lim_{\eta\rightarrow 0}\sum_{\mu=1}^{N_S}\frac1{z-\ve^s_\mu-i\eta} 
\ee
We use the power expansion ansatz for the inverse of the Cauchy transform to be
\be\label{eq:pert_ans}
{\tilde G}_{H_F}^{-1}(w) = {G}_{H_F}^{-1}(w)+\alpha A(w)+O(\alpha^2)
\ee
where $A(w)$ is a surface correction which can be obtained from the consistency condition
\be\label{eq:consisntency}
{\tilde G}_{H_F}\Bigl({\tilde G}_{H_F}^{-1}(w)\Bigl)=w
\ee
Inserting Eq.\eqref{eq:full_Greensf} and Eq. \eqref{eq:pert_ans} into Eq.\eqref{eq:consisntency}, one derives
\be
 A(w) =-\frac 1{\mathcal N} \frac {2\pi}{N_S}\sum_\mu\frac1{{G}_{H_F}^{-1}(w)-\ve_\mu-i\eta},\qquad \mathcal N = \frac{\partial G_{H_F}(z)}{\partial z}\Biggl|_{z=G^{-1}_{H_F}(w)} = \frac{1}{\Delta_0}(1+w^2)^{3/2}
\ee
Using Eq.\eqref{eq:GF_inverse_full} we calculate the Cauchy transform of the effective Hamiltonian to get
\be
\tilde G_{H_F^{\rm eff}}^{-1}(w)  = G_{H_F^{\rm eff}}^{-1}(w) + \alpha A(w)+O(\alpha^2).
\ee
Taking the functional inverse we use a power series expansion once more to get
\be
\tilde G_H(z) = G_H(z) + \alpha B(z) + O(\alpha^2),
\ee
where
\be
B(z) = -\frac 1{\mathcal N_2} A\Bigl(G_{H_F^{\rm eff}}(z)\Bigl) ,\qquad  \mathcal N_2 = \frac{\partial G^{-1}_{H_F^{\rm eff}}(w)}{\partial w}\Biggl|_{w=G_{H_F^{\rm eff}}(z)} = \lambda^2+\frac{\Delta_0}{(1+w^2)^{3/2}}
\ee
\be
B(z) = g\frac {2\pi}{N_S'}\sum_{\mu=1}^{N_S}\frac1{G_{H_F}^{-1}(G_H(z))-\ve_\mu-i0_+}, \qquad g = \frac{N_S}{1+\lambda^2\Delta_0^{-1}(1+G_{H_F^{\rm eff}}(z)^2)^{3/2}}.
\ee
For real $z$, if the condition $|z|<\Delta(\lambda)$ is satisfied, the prefactor $g$ is real, and the midgap states' energies $\ve'_\mu$ are defined by the new poles:
\be
\ve_\mu = G^{-1}_{H_F}(G_{H_F^{\rm eff}}(\ve'_\mu)).
\ee
This condition leads to the following solution
\be
\ve'_\mu = \ve_\mu\Biggl(1-\frac{\pi\rho_0\lambda^2}{\sqrt{\Delta_0^2-\ve_\mu^2}}\Biggl), \qquad \lambda<\lambda^\mu_{\rm c}
\ee
This expression describes continuous deformation of the surface states spectrum without opening the gap.
\section{Disorder effect on Floquet topological systems}
As discussed in the main text, disorder added to a Floquet topological can destroy the topological phase. 

To support the general analytical framework above, we now apply it to specific example to demonstrate the signatures of disorder-induced phase transition in finite size Floquet systems. We consider two non-interacting Floquet topological models: Kitaev chain [4] and Bernevig-Hughes-Zhang model [5]. In both cases, we study a time-periodic Hamiltonian in the form
\be\label{eq:driving_type}
H(t) = H_0+V\theta(t),\qquad 
\theta(t) = 
\begin{cases}
+1,\quad n\tau <t\leq n\tau + \tau/2,\\
-1,\quad n\tau + \tau/2 < t \leq (n+1)\tau,
\end{cases}
\ee
where $H_0$ describes disordered insulator with trivial band topology, $V$ is a local driving field, and $\tau$ is driving period. We perform our analysis by studying Floquet Hamiltonian, which is a functional of $H_0$ and $V$
\be\label{eq:Floquet_def}
H_F = H_F\{H_0,V\}  = \frac i\tau\log\Bigl(e^{-i(H_0-V)\tau/2}e^{-i(H_0+V)\tau/2}\Bigl)
\ee
We study the behavior of the gap in quasi-energy spectrum of $H_F$ and its topological invariants as a function of static disorder. Also, to justify our approximations, we study the structure of disorder corrections to the Floquet Hamiltonian.

\subsection{1D Example: Kitaev chain}
Kitaev chain is an example of 1D topological superconductor [4]. In terms of electron operators in Fock space, the Kitaev chain Hamiltonian and corresponding driving field can be written as
\be
\hat H_0 = (Jc^\dag_ic_{i+1}+{\rm h.c.})+(\mu+h_i) c^\dag_ic_i - (D c_ic_{i+1}+{\rm h.c.}),\qquad \hat V = fc^\dag_ic_i
\ee
where $c_i$ is an electron creation operator at site $i$, $J>0$ is hopping constant, $D>0$ is the superconducting gap, $\mu>0$ is chemical potential, and $h_i\in[-W,W]$ is on-site random disorder. For numerical study, it is convenient to consider the Bogoliubov-de Gennes (BdG) form of the Hamiltonian $H_0$ and driving field $V$ defined such that
\be
\hat H_0 = \hat C^\dag H_0 \hat C, \qquad \hat V = \hat C^\dag V \hat C,
\ee
where $\hat C = \{c_1\dots c_L,c_1^\dag\dots c_L^\dag\}$, and $L$ is size of the system.

As a result, the exact form of BdG Hamiltonian can be written as $2L\times 2L$ matrices
\be
H_0 = \sum_i\Bigl((J\sigma_z+iD\sigma_y)|i\>\<i+1|+{\rm h.c.}\Bigl)+(\mu+h_i)\sigma_z|i\>\<i|,\qquad V = \sum_i f\sigma_z|i\>\<i|
\ee
where $\sigma_i$ are $2\times2$ Pauli matrices. This expression can be compared to Eqs.(5)-(6) in the main text. 

For a particular choice of parameters corresponding to trivial static phase (we use $J=D=1$, $\mu=4.5$, and $f=1.5$), the drive system  exhibits several transitions at finite driving frequency with $0$-quasienergy or/and $\pi$-quasienergy Majorana states. We focus on stability of $0$-quasienergy Majorana state at driving period $\tau=1.1$ characterized by quasienergy gap $\Delta_0 \approx 0.12\, \tau^{-1}\ll \Omega$ and density of states $\rho_0 \approx 0.32\,\tau$. The density of states of resulting Floquet Hamiltonian $H_F$ Eq.\eqref{eq:Floquet_def} for the system size $L=10^2$ is shown in Fig. 3A in the main text (on the right). 

The topological invariance for Floquet Hamiltonian can be computed similar to the equilibrium Majorana fermion [6] by:
\be
Q = Q_0Q_\pi = {\rm sign}\Bigl({\rm Pf}(iH_F^m)\Bigl),\qquad H_F^m = U_mH_FU_m^\dag
\ee
where $Q_0$, $Q_\pi$ is topological charges for zero-quasienergy and $\pi$ Majorana states correspondingly, $H_F$ is Floquet Hamiltonian for the time dependent Hamiltonian with periodic boundary conditions, $U_m$ is a unitary transformation to the basis of Majorana fermions  converting $H_F$ into skew-symmetric matrix $H^m_F$, $I_c = \sum_i|i\>\<i|$ is the identity operator in coordinate basis, and $\rm Pf(\cdot)$ is a Pfaffian.

We suppose that disorder does not destroy $\pi$-quasienergy Majorana fermion which has much larger gap. Then, disorder induced transition for 0-quasienergy Majorana state can be characterized by the parameter
\be
N = -\<Q\>_{\rm dis}
\ee
If gapped topological phase $Q=-1$, thus $N=1$. In disordered gapless trivial phase the charge $Q=\pm 1$ with equal probability depending on disorder realization, i.e. $N = \<Q\>_{\rm dis} = 0$. The transition for the parameters chosen above is shown on Fig 2c in the main text.

Let us estimate the radius of convergence of the series for Floquet Hamiltonian in the case $\Delta=J$. For this, we use the criterion of convergence of Magnus expansion $\int_{0}^{\tau}||H(t)||dt\leq\pi$ [7]. Let us first focus on the system without disorder. The spectrum of the Kitaev chain is
\begin{equation}
\epsilon^i_k = \pm\sqrt{(\mu_i+J\cos k)^2+J^2\sin^2 k}
\end{equation}
where $k\in[-\pi,\pi]$ is the wavenumber, $i=1$ correspond to the instantaneous Hamiltonian during the first half of the step driving $n\tau<t<(n+\frac 12)\tau$, and $i=2$ is for the second part $(n+\frac 12)\tau<t<(n+1)\tau$, and $n$ is an integer.

Then we can derive that for any sign of $\mu_i$ the following holds
\begin{equation}
\int_{0}^{\tau}||H(t)||dt =  \frac\tau2\Bigl(\max_k|\epsilon^1_k|+\max_k|\epsilon^2_k|\Bigl) = \tau \bigl(J+\overline\mu\bigl),\qquad \overline\mu =\frac12(|\mu_1|+|\mu_2|). 
\end{equation}
 Therefore the convergence of Magnus expansion is guaranteed for $\tau<\tau_c$, where $\tau_c = \pi/(J+\overline\mu).$
 Adding disorder typically increases the bandwidth of the system. This result in a radius of convergence that is upper-bounded by that of the clean system. For the system shown in Fig. 3A in the main text, $J=\Delta = 1.0$, $f=1.5$, $\mu_1 = \mu-f = 3.0$ and $\mu_2 =\mu+f=6.0$, which gives the radius of convergence $\tau_c\approx 0.57$. 

\subsection{2D example: Bernevig-Hughes-Zhang model}

The Bernevig-Hughes-Zhang (BHZ) model is paradigmatic example of 2D topological insulator [5]. In this work we consider a discretized version of BHZ model on a square lattice of size $L_x \times L_y$ written as
\be\label{eq:BHZ_discrete}
H_0 = \sum_{x,y} \Bigl[\Bigl(\Gamma_x |x,y\>\<x+1,y|+\Gamma_y |x,y\>\<x,y+1|+{\rm h.c.}\Bigl)+ M_{x,y} |x,y\>\<x,y|\Bigl],\qquad V(t) = \sum_{x,y}f\sigma_z|x,y\>\<x,y|
\ee
where $x$ and $y$ are integers corresponding to the coordinates of the site on the square lattice. We choose $\Gamma_x$, $\Gamma_y$ and $M_{x,y}$ to be $2\times2$ matrices
\be
\Gamma_x = -i\frac{A}2 \sigma_x + B\sigma_z, \quad \Gamma_y = -i\frac{A}2 \sigma_y + B\sigma_z,\quad
M_{x,y} = h_{x,y}+(\mu-4B)\sigma_z
\ee
where $h_{x,y}\in[-W,W]$ is on-site scalar disorder. 


In the long wavelength limit the discrete Hamiltonian Eq.\eqref{eq:BHZ_discrete} reduces to the conventional form of BHZ Hamiltonian:
\be
H_0 = A\hat p_x\sigma_x+A\hat p_y\sigma_y+(\mu-B(\hat p_x^2+\hat p_y^2))\sigma_z + h(x,y), \qquad V= f\sigma_z
\ee
where $\hat p_x=-i\partial_x$ and $\hat p_y=-i\partial_y$ are continuous momentum operators, and  $h(x,y)$ is static disorder in continuous limit.

Similar to the Kitaev chain, at certain parameters describing non-topological case (we use $A=0.25$,$B=-0.25$, $\mu=1$) and resonant driving (we use $f=1$ and $\Omega = 2\pi/T = 0.4$), the driven system  exhibits a topological phase around $\pi$ quasienergy. DOS of the resulting Floquet Hamiltonian $H_F$ for system size $L_x=20$ (periodic b.c.), $L_y=20$ (open b.c.) is shown in Fig.$\,$2B in the main text (the color plot). The transition for the parameters chosen above is shown on Fig. (2)D in the main text.

To define the topological invariant, we use Bott index (BI) as topological invariant in the system. BI is used in number of previous works [8, 9] and is known to be equivalent to Chern number in presence of translational symmetry [10, 11]. Let us consider a two-band insulator. To define BI, first let us define a pair of unitary operators $U_1$ and $U_2$ represented by $L_x L_y\times L_x L_y$ matrices such that:
\be
\mathcal P_{E} e^{2\pi i X} \mathcal P_{E} = 
W\left(\begin{matrix}
0 & 0\\
0 & U_1
\end{matrix}\right)W^\dag,\qquad
\mathcal P_{E} e^{2\pi i Y} \mathcal P_{E} = 
W\left(\begin{matrix}
0 & 0\\
0 & U_2
\end{matrix}\right)W^\dag
\ee
where $\mathcal P_{E}$ is a projector on the lower band, operators $X = L_x^{-1}\sum_{x,y} x|x,y\>\<x,y|$ and $Y = L_y^{-1}\sum_{x,y} y|x,y\>\<x,y|$. The unitary transformation $W$ is chosen such that
\be
\mathcal P_{E}  = W\left(\begin{matrix}
0 & 0\\
0 & I
\end{matrix}\right)W^\dag
\ee

BI is defined as
\be
B = \frac 1{2\pi}{\rm Im}\Tr\log \Bigl(U_1U_2U_1^\dag U_2^\dag\Bigl)
\ee
The values of $B$ for each disorder realizations are integer even for finite system size. At the same time, disorder-averaged values for finite system can be an arbitrary real number. 

\subsection{Properties of Renormalized Disorder}

The driving-renormalized disorder $\delta V_F$ is defined as corrections to the Floquet Hamiltonian,
\be\label{eq:s_UF_formula}
\delta V_F = i\log U_F'-i\log U_F
\ee
where $\delta V$ is the original local disorder. A formal expansion can be used to represent $\delta V_F$ by 
\be\label{eq:Magnus_expansion_supp}
\delta V_F\equiv \sum_{\ell \ge 1} \delta V_\ell \text{ }\tau^{\ell-1},
\ee
where the terms in the expansion are
\be\label{eq:s_magnus_exp}
\delta V_1 = \delta V, \quad \delta V_\ell = \frac{1}{\tau^\ell}\Bigl[K_{\ell}\{H(t)+\delta V\} - K_{\ell}\{H(t)\}\Bigl],
\ee
and $K_\ell$ are the Magnus expansion coefficients. First three coefficients are
\begin{align}
&K_1\{A(t)\} = \int_0^\tau dt A(t),\qquad K_2\{A(t)\} = \frac12\int_0^\tau dt_1\int_0^{t_1} dt_2 [A(t_1),A(t_2)],\\
&K_3\{A(t)\} = \frac16\int_0^\tau dt_1\int_0^{t_1} dt_2\int_0^{t_2} dt_3 \Bigl([A(t_1),[A(t_2),A(t_3)]]+[A(t_3),[A(t_2),A(t_1)]]\Bigl).
\end{align}
At large driving frequency, renormalization is weak and $\delta V_F$ is close to the original disorder $\delta V$. 

To demonstrate this we visualize the statistics of level spacings of the spectrum of $\delta V_F$ defined by $\text{Spec}\left(\delta V_F\right) \equiv \{v_n\}$, where $v_1\le v_2\le\dots\le v_{max}$. The consecutive level spacings are defined by
\be
s_n \equiv v_{n+1}-v_n.
\ee
We focus on level spacings in $\ve$-vicinity of the center of the spectrum, namely $\overline v-\ve S<s_n<\overline v+\ve S$, where $S = v_ {max}-v_1$.

We compare the distribution of ${s_n}$ to GOE and GUE defined by
\be
{\rm GOE:} \qquad p_1(s) = \frac {\pi}2 s\exp(-\frac\pi 4s^2),\qquad
{\rm GUE:} \qquad p_2(s) = \frac {32}{\pi^2} s^2\exp(-\frac4\pi s^2)
\ee
The level statistics of the renormalized disorder operator $\delta V_F$ in the vicinity of $\ve=0.1$ to the center is shown on Fig. 2A. Notably, the spacing of the disorder corrections for Kitaev chain is closer to GOE, while BHZ disorder correction is better described by GUE. 

As expected, the effect of $\delta V_F$ in spectral properties of the Floquet is similar to GRM. Hence, it may not lead to Anderson localization in low dimensions. 

We, however, do not believe that $\delta V_F$ can always be replaced by GRM. For example, $\delta V_F$ must respect causality and Lieb-Robinson bounds characterized by velocity $v_{\rm LR} \sim T$. This property cannot be captured by a GRM. Although, we believe that spectral properties of matrix $H_F+\delta V_F$ can be accurately described if we replace $\delta V_F$ by GRM. As can be seen the numerical results give good agreements to the analytical formulas.

Lastly, we derive equations that quantify the dependence of the {\it effective} disorder strength $\lambda$ on the period $\tau$ and the disorder strength $W$. Let us consider the particular form of driving in Eq.$\,$\eqref{eq:driving_type} and apply Eq.$\,$\eqref{eq:s_magnus_exp} to derive the Floquet disorder correction. To the lowest orders we have
\be
\delta V_F = \delta V -\frac\tau4 [\delta V,V]+\frac{\tau^2}{24}[\delta V,[V,H_0]] +O(\min(W^2 \tau^3, W\tau^4))
\ee
Assuming that $\delta V = \sum_{\vec r}h_{\vec r}|\vec r\>\<\vec r|$, where $h_{\vec r}\in[-W,W]$, the expression for the effective disorder strength becomes
\be\label{eq:alphTau}
\lambda = \sqrt{\phi(\delta V_F^2)} = \alpha(\tau)\frac W{\sqrt{3}}+O(W^2),
\ee
where we used that the disorder is  uncorrelated at different lattice sites giving $\sum_{\vec r,\vec r'} h_{\vec r}h_{\vec r'}= \frac{W^2}3\delta_{\vec r,\vec r'}$. The parameter $\alpha(\tau)$ has the form 
\be
\alpha(\tau) = 1+\tau^2\Delta (V_\tau), \qquad V_\tau = V-\frac{\tau}{6}[V,H_0]+O(\tau^2)
\ee
where $\Delta(V_\tau) = \sum_{\vec r}\Bigl(\<\vec r|V_\tau|\vec r \>^2-\<\vec r|V_\tau^2|\vec r\>\Bigl)$. If in addition the applied periodic filed is local, $\Delta(V)=0$, the finite frequency have forth order corrections in period, $\alpha(\tau) = 1+O(\tau^4)$.
More general result can be obtained for arbitrary period $\tau$ using the expansion over small disorder strength,
\be
e^{-i(H_1+\delta V)\tau/2} = e^{-iH_1\tau/2} \Bigl(1-i\hat F_1\delta V\Bigl)+O(\delta V^2), \qquad e^{-i(H_2+\delta V)\tau/2} =  \Bigl(1-i\hat F_2\delta V\Bigl)e^{-iH_2\tau/2}+O(\delta V^2)
\ee
where we define the superoperator
\be
\hat F_{i} = \frac{1-e^{-\hat G_{H_i}}}{\hat G_{H_i}} = \sum_k \frac{(-1)^k}{(k+1)!}\hat G_{H_n}^k,
\qquad \hat G_{H}B = i\frac \tau 2[H,B].
\ee
Then, the Floquet operator for disordered system can be expressed as a perturbed operator for the clean system,
\be
U_F' = U_F-\frac{i\tau}2 e^{-iH_1\tau/2}\bigl(\hat F_1\delta V+\hat F_2\delta V\bigl)e^{-iH_2\tau/2}+O(\delta V^2)
\ee
The expression for Floquet disorder corrections can be derived from Eq.$\,$\eqref{eq:s_UF_formula} using Taylor expansion we the derivative of the matrix logarithm
\be
\frac{d\log A}{dt} = \frac1{2\pi i}\int dz \log(z)\frac1{z-A}\frac{dA}{dt}\frac1{z-A}
\ee
As a result, the expression for Floquet disordered correction in the limit of weak disorder reads
\be
\delta V_F = \mathcal F\delta V+O(\delta V^2),\quad {\rm where}\quad  \mathcal F\delta V=  \frac 1{4\pi i}\int dz\log(z)\frac1{z-U_F}e^{-iH_1\tau/2}\bigl(\hat F_1+\hat F_2\bigl)\delta Ve^{-iH_2\tau/2}\frac 1{z-U_F}.
\ee
This gives us the dependence of $\alpha$ on $\tau$:
\be
 \alpha(\tau) = \sum_{\vec r}\sqrt{3\phi\Bigl((\mathcal F |\vec r \>\<\vec r|)^2\Bigl)}
\ee
Eq. \eqref{eq:alphTau} is valid for static disorder strengths that are smaller than any other relevant energy in static Hamiltonian such as next-neighbour hopping parameter.

If the driving frequency is finite, the series \eqref{eq:driving_type} can be divergent and $\delta V_F$ may be essentially different from $\delta V$. In particular, in many problems $\delta V$ is simply a diagonal random matrix. However, low-energy $\delta V_F$ can have properties of generic random matrix (GRM) as argued in the paper (see for example Fig.$\,$2) and further elaborated on below.
\vspace{3em}
\begin{small}
\begin{enumerate}[label={[\arabic*]}]
\item A. Nica and R. Speicher, {\it Lectures on the combinatorics of free probability}, Vol. 13 (Cambridge University Press, 2006).

\item  R. S. James A. Mingo, {\it Free Probability and Random Matrices} (Springer New York, 2017).
\item  R. Movassagh and A. Edelman, arXiv preprint arXiv:1710.09400  (2017).

\item A. Y. Kitaev, Phys. Usp. \textbf{44}, 131 (2001).

\item B. A. Bernevig, T. L. Hughes,  and S.-C. Zhang, Science \textbf{314}, 1757 (2006).

\item L. Jiang, T. Kitagawa, J. Alicea, A. R. Akhmerov, D. Pekker, G. Refael, J. I. Cirac, E. Demler, M. D. Lukin,  and P. Zoller, Phys. Rev. Lett. \textbf{106}, 220402 (2011).

\item S. Blanes, F. Casas, J. Oteo,  and J. Ros, Phys. Rep. \textbf{470}, 151 (2009).

\item P. Titum, N. H. Lindner, M. C. Rechtsman,  and G. Refael, Phys. Rev. Lett. \textbf{114}, 056801 (2015).

\item R. Ducatez and F. Huveneers, Ann. Henri Poincaree \textbf{18}, 2415 (2017).

\item  T. A. Loring and M. B. Hastings, Europhys. Lett. \textbf{92}, 67004 (2011).

\item D. Toniolo, arXiv preprint arXiv:1708.05912  (2017).
\end{enumerate}
\end{small}
\end{document}